\documentclass[amsmath,amssymb,aps,prb,twocolumn,floatfix,10pt,superscriptaddress]{revtex4-2}

\usepackage{mathtools}
\usepackage[caption = false]{subfig}
\usepackage{graphicx}
\usepackage{standalone}
\usepackage{dcolumn}
\usepackage{bm}
\usepackage{xcolor}
\usepackage{physics}
\usepackage{enumitem}
\usepackage{calligra}
\usepackage[colorlinks = true,linkcolor = red,citecolor = magenta]{hyperref}
\usepackage{natbib}
\usepackage{orcidlink}
\usepackage[normalem]{ulem}
\usepackage{pdfpages}

\makeatletter
\AtBeginDocument{\let\LS@rot\@undefined}
\makeatother

\DeclareGraphicsExtensions{.pdf,.eps,.png,.jpg,.mps}

\newcommand{\pcsadd}{Center for Theoretical Physics of Complex Systems, Institute for Basic Science (IBS), Daejeon, Korea, 34126}
\newcommand{\trappedions}{Center for Trapped Ions Quantum Science, Institute for Basic Science, Daejeon 34126, Republic of Korea}
\newcommand{\ustadd}{Basic Science Program, Korea University of Science and Technology (UST), Daejeon 34113, Republic of Korea}
\newcommand{\nziasadd}{Centre for Theoretical Chemistry and Physics, The New Zealand Institute for Advanced
Study (NZIAS), Massey University Albany, Auckland 0745,
New Zealand}

\makeatletter
\renewcommand*{\fnum@figure}{{\normalfont\bfseries \figurename~\thefigure}}
\renewcommand*{\@caption@fignum@sep}{\textbf{:}}
\makeatother

\newcommand{\bea}{\begin{eqnarray}}
\newcommand{\eea}{\end{eqnarray}}

\newcommand{\appropto}{\mathrel{\vcenter{
  \offinterlineskip\halign{\hfil$##$\cr
    \propto\cr\noalign{\kern2pt}\sim\cr\noalign{\kern-2pt}}}}}

\newlength\mylen
\newlist{mycases}{enumerate}{1}
\setlist[mycases,1]{label=\textbf{Case~\arabic*.}, 
  labelwidth=\dimexpr-\mylen-\labelsep\relax,leftmargin=0pt,align=right}

\begin{document}

\title{
Disorder-Tailored Delocalization
}

\author{Yeongjun Kim}
    \email{yeongjun.kim.04@gmail.com}
        \affiliation{\pcsadd}
        \affiliation{\trappedions}

\author{Supriyo Ghosh }
    \email{supriyoghosh711@gmail.com}
    \affiliation{Department of Physics, Indian Institute of Technology-Kanpur, Kanpur 208016, India}

\author{Sergej Flach\,\orcidlink{}}
    \email{sflach@ibs.re.kr}
    \affiliation{\pcsadd}
    \affiliation{\trappedions}
    \affiliation{\ustadd}
    \affiliation{\nziasadd}

\date{\today}

\begin{abstract}

We derive disorder fields tailored by the details of a choice of a delocalized wave function.
We first investigate the unidirectional Hatano-Nelson chain and its localization properties under $M$-base diagonal disorder with variable weights.
The spectrum forms loops in the complex plane and the loop parameter is a good quantum number similar to a momentum.
All eigenstates are subexponentially `localized', i.e. the logarithm of the absolute value of the wave function performs a random walk in space, and are characterized by a corresponding length scale $\xi_{sel}$ as shown in 1998 by Silvestrov for the general Hatano-Nelson chain.
For $M=2$ real-valued binary disorder with equal weights the model was solved in Zeitschrift für Naturforschung A 81 421, yielding Cassini oval spectral loops and a diverging subexponential localization length for two eigenstates and for disorder weaker than a critical value set by the hopping strength.
When the disorder field for any $M$ and arbitrary weights is confined to circles in the complex plane with radius equal to the hopping strength, the circle center will belong to the spectrum and to one of the spectral loops, and host a plane-wave-like eigenstate with diverging $\xi_{sel}$.
We generalize to tailoring on-site disorder for a given eigenstate at a given energy, for any lattice dimension, and any hopping field (both ordered and disordered), for Hermitian and non-Hermitian systems.
We exemplify by constructing a one-dimensional chain with Anderson-localized eigenstates hosting a completely delocalized one with random phases. 
The localization length diverges as $1/|E|^{2/3}$ 
upon approaching the delocalized state.
Our method can be used for the systematic construction of disorder fields which host predefined eigenstates with arbitrary properties. 

\end{abstract}

\maketitle

\section{Introduction}
\label{sec:introduction}

Wave equations in disorder fields are known to produce localization~\cite{Anderson1958PR}.
Symmetries may protect certain eigenenergies from that fate such that their eigenstates remain delocalized.
One example is the chiral symmetry of a tight-binding chain which keeps a delocalized state at the chiral energy point for random hopping fields~\cite{PhysRevB.49.3190,balents1997delocalization}. 
While states are exponentially localized for energies away from the chiral point, the localization length will diverge logarithmically at the chiral point~\cite{theodorou1976extended,fleishman1977fluctuations,soukoulis1981off}. 
The delocalized state will show subexponential fluctuations as the logarithm of the absolute value of the wave function will perform a random walk in space~\cite{PhysRevB.49.3190,balents1997delocalization}. 
Another example is a chain of random masses connected by harmonic springs which hosts a completely uniform delocalized state at zero frequency due to Galilean symmetry. 
Low frequency states will show a power law divergence of their respective exponential localization length as $1/\omega^2$~\cite{matsuda1970localization,ishii1973localization}. 
Different symmetries appear to enforce different scaling laws of approaching delocalized states, as well as the nature of these delocalized states.

In this work we will systematically derive non-Hermitian and Hermitian disordered lattice models which host delocalized states and may serve as vehicles for further localization-delocalization transition studies and their classifications. 
We will use an inverse tailoring method (see also Refs.~\cite{rodriguez2012controlled,makris2017wave} for related attempts). 
We will start with the unidirectional Hatano-Nelson chain with binary disorder which was recently analyzed in Ref.~\cite{ghosh2026spectral}. 
There a complete delocalization of the in general subexponentially fluctuating eigenfunctions was observed for specific eigenvalues. 
Here we show that this is in fact a simple consequence of disorder tailoring. 
For any choice of the two disorder potential values which can be located anywhere in the complex plane we show that if a circle with unit radius (in units of the hopping)  can run through both complex numbers, then the center of the circle must belong to the spectrum and host a completely delocalized eigenstate. 
We will generalize these results to arbitrary weights of the binary disorder, then generalize to $M$-base disorder, and finally arrive at disorder tailoring of almost any type of system.

\section{Disordered Hatano-Nelson models}
\label{sec:disordered_hatano_nelson_models}

\subsection{General remarks}
\label{sec:general_remarks}

Transverse magnetic fields can lead to vortex depinning from columnar defects in superconducting materials. 
The resulting vortex dynamics turns non-reciprocal and can be modeled by non-Hermitian disordered Hamiltonians as shown by Hatano and Nelson~\cite{HN1996PRL,hatano1997vortex,PhysRevB.58.8384}, and later also by Silvestrov~\cite{silvestrov1999vortices}. 
Similar results were obtained by Nelson and Shnerb when starting from a completely different problem of active matter~\cite{nelson1998non}. 
For the simplest one-dimensional Hatano-Nelson (HN) chain the presence of disorder results in a replacement of Anderson localization by subexponential localization $\exp(-\sqrt{x/\xi_{sel}})$ with a characteristic length $\xi_{sel}$ as shown by Silvestrov~\cite{PhysRevB.58.R10111}. 

The limiting case of the unidirectional Hatano-Nelson (uHN) chain is the irreducible non-Hermitian building block of the chiral bipartite Hermitian Su-Schrieffer-Heeger (SSH) model which was first introduced to describe the electrical conductivity of polyacetylene chains~\cite{SSH1979PRL}. 
In a recent work it was shown that the uHN chain with diagonal equal probability binary disorder allows for states with diverging subexponential localization length $\xi_{sel}\rightarrow \infty$ which is related to spectral topology properties~\cite{ghosh2026spectral}. 
The spectrum forms Cassini ovals~\cite{cassini1730origine,basset1901elementary,cohen1962leibniz,lawden1999families,lawrence2013catalog} with one or two loops in the complex plane~\cite{feinberg1999non}. 
In the present work we generalize these results to base-$M$ disorder with arbitrary probability weights. 
The spectrum forms generalized Cassini ovals with up to $M$ separate loops in the complex plane. 
If the disorder values arrange on a circle with radius $t$ (the hopping strength) then the center of the circle belongs to the spectrum (and to one of the loops) and its eigenstate is completely delocalized with a diverging localization length $\xi_{sel}$.

\subsection{Unidirectional case}
\label{sec:unidirectional_case}

We consider the unidirectional Hatano-Nelson model consisting of \(N\) sites described by the non-Hermitian Hamiltonian:
\begin{eqnarray}
    H = \sum_{n=1}^{N} \epsilon_n \ a_{n}^{\dagger}a_{n} -t \ a_{n}^{\dagger}a_{n+1} 
    \label{eq:ham1}
\end{eqnarray}
\noindent where \(a_{n}\) (\(a_{n}^{\dagger}\)) is the annihilation (creation) operator at site \(n\). 
Here, \(\epsilon_n\) denotes the onsite potential, while \(t\) represents the unidirectional hopping amplitude from site \(n+1\) to site \(n\). 
We assume periodic boundary conditions (PBC) such that \(a_{N+1} \equiv a_{1}\). 
Throughout most of the work we will set $t=1$. This Hamiltonian~\eqref{eq:ham1} represents a unidirectional Hatano-Nelson~\cite{HN1996PRL} chain (uHN) with random onsite potential $\epsilon_n$ and constitutes the irreducible block of the Su-Schrieffer-Heeger (SSH)~\cite{SSH1979PRL} model with randomized intercell hoppings \cite{ghosh2026spectral}. 

We will consider right eigenstates of the uHN Hamiltonian \eqref{eq:ham1} as:
\begin{eqnarray}
    \vert \Psi \rangle = \sum_{n = 1}^{N} \psi_{n} \ a_{n}^{\dagger} \ \vert 0 \rangle .
    \label{eq:gen_eig_state}
\end{eqnarray}
The Schr\"odinger equation \(H\vert \Psi \rangle = E\vert \Psi \rangle\) yields the following set of coupled equations:
\begin{eqnarray}
    E\psi_{n} & = & \epsilon_n \ \psi_{n} -  t \ \psi_{n+1} \ ;\hspace{2mm} n = 1,2 \dots (N-1), \nonumber \\
    E\psi_{N} & = & \epsilon_N \ \psi_{N} -  t \ \psi_{1}.
    \label{eq:set_schr}
\end{eqnarray}
Let us rewrite the above equation as
\begin{equation}
    \frac{\psi_{n+1}}{\psi_n} = \frac{\epsilon_n -E}{t} \;. 
    \label{eq:TM1} 
\end{equation}
It follows
\begin{eqnarray}
    \prod_{n=1}^{N} \frac{\psi_{n+1}}{\psi_{n}} = \prod_{n=1}^{N}\frac{\epsilon_n-E}{t} = 1. \label{eq:TM2}
\end{eqnarray}
This is a remarkably simple set of equations. 
Equation~\eqref{eq:TM1} tells that any given disorder realization in the potential $\epsilon_n$ is imprinted straight into the ratio of the wave function amplitudes on neighbouring sites for a given eigenstate with energy $E$. 
If there is no correlation in the disorder, there is none in the wave function ratios. 
Therefore and because of the unidirectionality of the hopping, there is no room for destructive interference. 
Solutions of Eq.~\eqref{eq:TM2} result in the set of eigenenergies $E$. 
These equations are directly related to the zero energy eigenstates of the chiral bipartite Hermitian Anderson chain (HAC) eigenvalue problem with random hopping $E \psi_n =-t_{n,n+1}\psi_{n+1} - t_{n-1,n}\psi_{n-1}$ which reads $\frac{\psi_{n+1}}{\psi_{n-1}} = -\frac{t_{n,n-1}}{t_{n+1,n}}$. 
The HAC model shows exponential Anderson localization $\psi_n \sim \exp{-n/\xi}$ with finite exponential localization length $\xi$ at nonzero energies, and while the localization length diverges at $E=0$, the two eigenstates at $E=0$ show subexponential fluctuations with the log of the wave function performing a random walk: $\psi_n \sim \exp{\pm \sqrt{n/\xi_{sel}}}$ with a finite subexponential localization length $\xi_{sel}$~\cite{PhysRevB.49.3190}.

\subsection{Clean system}
\label{sec:clean_system}

For the clean system $\epsilon_n=C$ and $t=1$ the ansatz \(\psi_{n} = A \ e^{iqn}\) results in the dispersion relation \(E = C - e^{iq}\).
The boundary conditions restrict the allowed values of the wavevector to \(q = \frac{2\pi s}{N}\), where \(s = 0,1,\dots (N-1)\). 
In the complex energy plane, the eigenvalue spectrum of the ordered uHN model forms a circle centered at $C$ with radius $1$. The corresponding eigenvectors are extended plane waves, where the wavevector $q$ corresponds to the angular position of a point on the spectral circle.

\subsection{Subexponential localization}
\label{sec:subexponential_localization}

Consider some general disorder in $\epsilon_n$. Then it follows from \eqref{eq:TM1} that
\begin{equation}
    \ln(|\psi_{n+1}|) = \ln(|\psi_n|) + \ln (\left| \frac{\epsilon_n -E}{t} \right| ) \;.
    \label{eq:TMLOG}
\end{equation}
This is a random walk for the logarithm of the absolute value of the wave function. 
Any potential bias due to a nonvanishing first moment of the random term in the rhs of \eqref{eq:TMLOG} will be cancelled by the boundary conditions. 
Consequently a typical eigenstate of the disordered uHN chain will show subexponential localization $\psi_n \sim {\rm e}^{-\sqrt{n/\xi_{sel}}}$ with the inverse of the localization length $\xi_{sel}$ equal to the variance (since the standard deviation of the random walk is $\sigma \sqrt{n} = \sqrt{n/\xi_{sel}}$) of the logarithmic disorder field $\ln (\left| \frac{\epsilon_n -E}{t} \right| )$. 
Our subsequent goal is to figure constraints on the disorder field $\epsilon_n$ which allow for at least one eigenstate in the uHN spectrum to be completely delocalized with $1/\xi_{sel}=0$.  

\subsection{Complete delocalization in disordered uHN}
\label{sec:complete_delocalization_in_disordered_uHN}
Let us first state the general condition for a delocalized eigenstate in uHN.
From Eq.~\eqref{eq:TM1}, the condition for constant amplitude ratio is given as
\begin{align}
    |\epsilon_n - E_0| = |t|
    \label{eq:condition1}
\end{align}
for some \(E_0\).
Geometrically, \(\epsilon_n\) have to be located on a circle of radius \(|t|\), centered at \(E_0\).
We may write \(\psi_{n+1}/\psi_n = e^{i\Delta\phi_n}\).
Then the periodic boundary condition Eq.~\eqref{eq:TM2} gives:
\begin{align}
    \sum \Delta\phi_n = 2\pi m.
    \label{eq:condition2}
\end{align}
When Eq.~\eqref{eq:condition1}-\eqref{eq:condition2} are satisfied, uHN+PBC has a completely delocalized state. 
Since we can always gauge the complex energy spectrum by parallel shifts anywhere in the complex plane, what matters are the phases $\Delta \phi_n$ of $\epsilon_n$ on the circle. 
For a finite system, the accumulated phase mismatch modulo $2\pi$ is generically of $O(1)$. 
The corresponding displacement of the nearest eigenvalue from $E = 0$ is of $O(1/N)$.

For finite system size $N$ each spectral loop (see Secs.~\ref{sec:binary_disorder} and~\ref{sec:base_M_disorder}) will be discretized by a set of points whose number is proportional to $N$. 
These points will be at an average distance $1/N$ on the loop. 
Unless we further finetune the disorder by choosing a subset of the disorder realization $\{\epsilon_n\}$ such that Eq.~\eqref{eq:condition2} is exactly satisfied, the closest eigenvalue on the loop passing through the circle center will be at a distance $\sim 1/N$ from the designated circle center. 
Its eigenvector will therefore be almost but not completely delocalized for large but finite $N$. 
For $N \rightarrow \infty$ complete delocalization will be obtained.

\subsection{Binary disorder}
\label{sec:binary_disorder}

Consider binary disorder with $\epsilon_n$ taking two complex values $h_1$ and $h_2$.
For a finite chain, we denote by $N_1$ and $N_2$ the numbers of sites with $\epsilon_n=h_1$ and $\epsilon_n=h_2$, respectively, and define the fractions $p_1=N_1/N$ and $p_2=N_2/N$, with $p_1+p_2=1$.

\subsubsection{Equal weights}
\label{sec:equal_weights}

Let us recap the case $p_1=p_2=1/2$ with real values $h_1=-h_2\equiv h$ which was studied in \cite{ghosh2026spectral}. 
The eigenvalue problem is reduced to
\begin{eqnarray}
    (E-h)(E+h)=t^2{\rm e}^{iq}\;,\; 0 \leq q \leq 2\pi. \nonumber
\end{eqnarray}
Since $|E-h|\; |E+h| =|t|^2$ it follows that we are searching for allowed values of $E$ in the complex plane such that the product of their distance to two points $\pm h$ are constant. 
These curves are loops and known as Cassini ovals~\cite{cassini1730origine,basset1901elementary,cohen1962leibniz,lawden1999families,lawrence2013catalog} with one or two loops in the complex plane encircling $\pm h$ \cite{feinberg1999non}:
\begin{eqnarray}
    E_{\pm} = \pm \sqrt{h^2 + t^2 \ e^{iq}}.
\end{eqnarray}
In polar form, the eigenvalues can be expressed as \cite{ghosh2026spectral}:
\begin{eqnarray}
    E_{\pm} = \pm \sqrt{R} \ e^{i\phi/2},
\end{eqnarray}
where
\begin{gather}
    \notag R = \sqrt{h^4 + t^4 + 2h^2t^2\cos(q)},\\
    \phi=\arg\left(h^2+t^2e^{iq}\right). \nonumber
\end{gather}
The eigenstates will be in general subexponentially localized and be characterized by a finite localization length $\xi_{sel} =\frac{1}{\left[\ln\left|(E+h)/t\right|\right]^2}$ and participation number $P = \frac{\left( \sum_{n=1}^{N} |\psi_{n}|^2 \right)^2}{\sum_{n=1}^{N} |\psi_{n}|^4}$~\cite{ghosh2026spectral}. 

A surprising consequent observation for the uHN chain with binary disorder is the following: if we can draw a circle with radius $t=1$ in the complex plane through $\pm h$ then the center of the circle is an allowed eigenvalue, and its corresponding eigenstate is completely and strictly delocalized with $1/\xi_{sel}=0$, $1/P=1/N$ and $|\psi_{n+1}|/|\psi_n|=1$ for all $n$. 
These states exist as long as $h\leq 1$ to satisfy the circle condition~\eqref{eq:condition1}, $|h-E| = |-h-E| = 1$. For $|h|<1$, we can draw two unit circles passing through $\pm h$, with centers at the two points where the Cassini oval intersects the imaginary axis, $E_\pm=\pm i\sqrt{1-h^2}$, resulting in two completely delocalized states. 
For $h=1$ the two centers merge at the origin, which hosts a single completely delocalized eigenstate when the PBC condition~\eqref{eq:condition2} is satisfied.
These states were first reported and analyzed in \cite{ghosh2026spectral}. With $\psi_n=|\psi_n| {\rm e}^{i\phi_n}$ the phases $\phi_{n+1} = \phi_n + \arg (\epsilon_n-E)$~\eqref{eq:TM1} carry the imprint of the chosen disorder field realization.

It is straightforward to generalize the above results to binary disorder with $h_1$ and $h_2$ being any choice of complex numbers. 
The spectrum will be again given by Cassini ovals. 
As long as the distance $|h_1 - h_2| \leq 2t$ the centers of the above two circles of radius $t=1$ running through both points will be eigenvalues with completely delocalized eigenstates.

\subsubsection{General weights}
\label{sec:general_weights}

If $p_1\neq 1/2$ most of the results in Sec.~\ref{sec:equal_weights} carry over. The spectrum will be given by generalized asymmetric Cassini ovals (see also \cite{longhi2021spectral}). 
The existence of completely delocalized states is actually even more generic as it is again exactly given by the possibility to draw circles of radius $t=1$  through both $h_{1,2}$ irrespective of their weights in the disorder realization. 
The centers of the circles are then their allowed eigenvalues. 

\begin{figure}[htbp]
\centering
    \includegraphics[width = 0.99\linewidth]{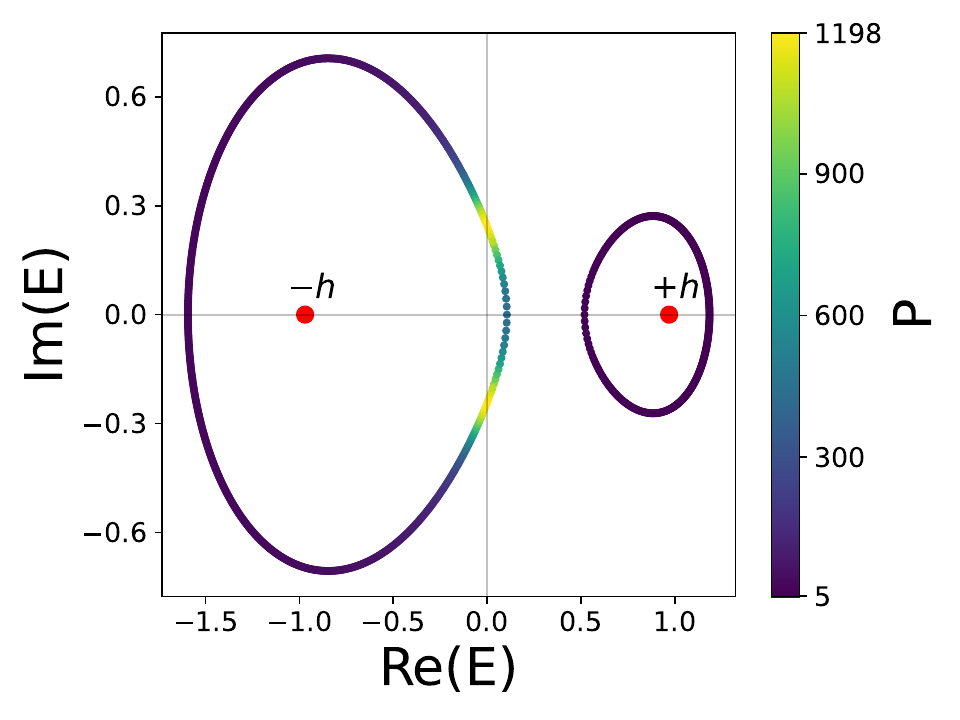}
    \caption{
        Complex energy spectra of the binary disordered uHN chain with probabilities $p_1=1/3$ ($+h$) and $p_2=2/3$ ($-h$). 
        Parameters: $t = 1, h=0.97$. The left Cassini loop cuts through the imaginary axis in two points, which are the precise locations of the delocalized states. 
        The participation number values $P$ are encoded in the color scheme. System size $N=1200$. 
        \label{fig:binary_disorder_eigvals}
    }
\end{figure}

Let us choose real values $h_1=h$ and $h_2=-h$ and $p_1=1/3$ and $p_2=2/3$.
The eigenvalue problem is reduced to 
\[
    (h-E)(E+h)^2 = {\rm e}^{iq}, 0 \leq q \leq 2\pi.   
\]
The spectrum is invariant under reflections at the real axis (while it is not under reflections at the imaginary one). 
Therefore the spectral bifurcation from one generalized Cassini loop to two loops happens when the two loops touch on the real axis. 
A simple calculation results in the bifurcation value $h_b=(27/32)^{1/3} \approx 0.9449$. For $h_b < h < 1$, we can still draw two circles and find two completely delocalized eigenstates.
The centers of the two circles will be now hosted by one of the two separated (asymmetric) Cassini loops. 
Therefore the previously observed correlation between spectral topology changes and the existence of completely delocalized eigenstates for $p_1=p_2=1/2$ in \cite{ghosh2026spectral} is accidental and does not hold for the general case of arbitrary weights. 
We plot the above spectrum for $h=0.97$ and $N=1200$ as obtained from exact diagonalization in Fig.~\ref{fig:binary_disorder_eigvals} together with the participation number $P$ in color code. 

\subsection{Base-\texorpdfstring{$M$}{M} disorder}
\label{sec:base_M_disorder}

Consider base-$M$ disorder with $\epsilon_n$ taking $M$ different values $h_1,h_2,...,h_M$ in the complex plane. 
We assume here that the field is uncorrelated, but stress that the main results do not depend on this assumption. 

With \eqref{eq:TM1} and \eqref{eq:TM2} it follows that if the $M$ complex numbers are located on one circle with radius $t=1$ then the center of that circle belongs to the eigenvalue spectrum and hosts a completely delocalized eigenstate. 
Note that for $M\geq 3$ we can only find one such circle at most. The above result holds for any disorder realization, and for any distribution of relative weights between the $M$ values $h_{1,2,...,M}$.

Let us consider equal weights $p_m=1/M$. Then the eigenvalue spectrum is given by
\begin{equation}
    \prod_{m=1}^M (h_m - E) = {\rm e}^{iq}\;,\; 0 \leq q \leq 2 \pi .
    \label{eq:baseM_equal_weight_spectrum}
\end{equation}
It forms a generalized Cassini oval structure with at least one and at most $M$ loops in the complex plane.

\begin{figure}[htbp]
    \centering
    \includegraphics[width = 0.99\linewidth]{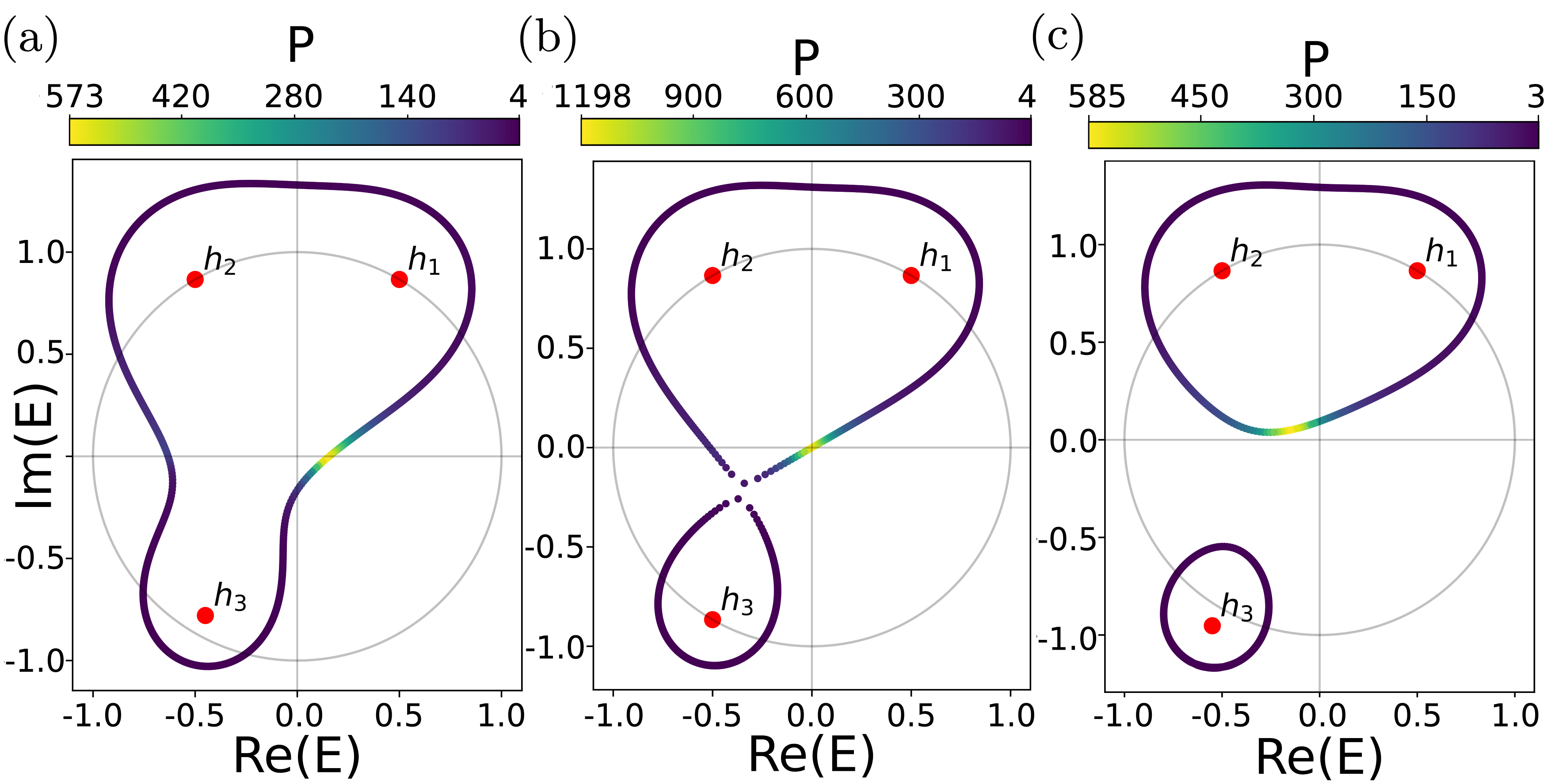}
    \caption{
        Complex energy spectra of the ternary disordered uHN chain. 
        The value of $P$ is shown by color code. 
        The unit circle is plotted with a black dotted line. 
        Parameter values: $t=1$, $N=1200$. $h_1 = e^{i\pi/3}$ and $h_2 = e^{2i\pi/3}$ are located on the unit circle. 
        (a) The spectrum consists of one Cassini loop, ~$h_3 = 0.9 e^{4i\pi/3}$ is slightly inside the unit circle, the energy $E=0$ does not belong to the loop, and the largest $P$ value reaches only $\approx 600=N/2$. 
        (b) The spectrum already consists of two Cassini loops,  ~$h_3 =  e^{4i\pi/3}$ is located precisely on the unit circle, the energy $E=0$ belongs to the top loop, and its largest $P$ value reaches $\approx 1200=N$. 
        (c) The spectrum continues to consist of two Cassini loops after $h_3 = 1.1 e^{4i\pi/3}$ is moved slightly outside of the unit circle. 
        The energy $E=0$ does not belong to any of the two loops, and the largest $P$ value dropped again to $\approx 600=N/2$.
    }
    \label{fig:ternary_disorder_eigvals}
\end{figure}

For the ternary case $M=3$ and the choice $h_1={\rm e}^{i\pi/3}$, $h_2={\rm e}^{i2\pi/3}$, $h_3={\rm e}^{i4\pi/3}$ all $h_m$ are located on a circle with radius $t=1$ and center at the origin. 
In the central plot in Fig.~\ref{fig:ternary_disorder_eigvals}(b) we show the eigenvalue spectrum. 
It forms a generalized Cassini oval structure of two loops. 
The top loop crosses right through the origin as predicted, where the eigenstate is completely delocalized.
In the right plot in Fig.~\ref{fig:ternary_disorder_eigvals}(c) we show the eigenvalue spectrum after moving $h_3$ slightly off the unit circle by choosing $h_3=1.1 {\rm e}^{i4\pi/3}$. 
Now the three disorder numbers cannot be arranged to be on a circle of radius $t=1$. 
Consequently the top loop detaches from the origin, and the completely delocalized eigenstate is lost. 
Likewise in the left plot in Fig.~\ref{fig:ternary_disorder_eigvals}(a) we move $h_3$ slightly off the unit circle by choosing $h_3=0.9 {\rm e}^{i4\pi/3}$. 
Both Cassini loops merged into one, and again the completely delocalized state is lost.

\begin{figure}[htbp]
    \centering
    \includegraphics[width = 0.99\linewidth]{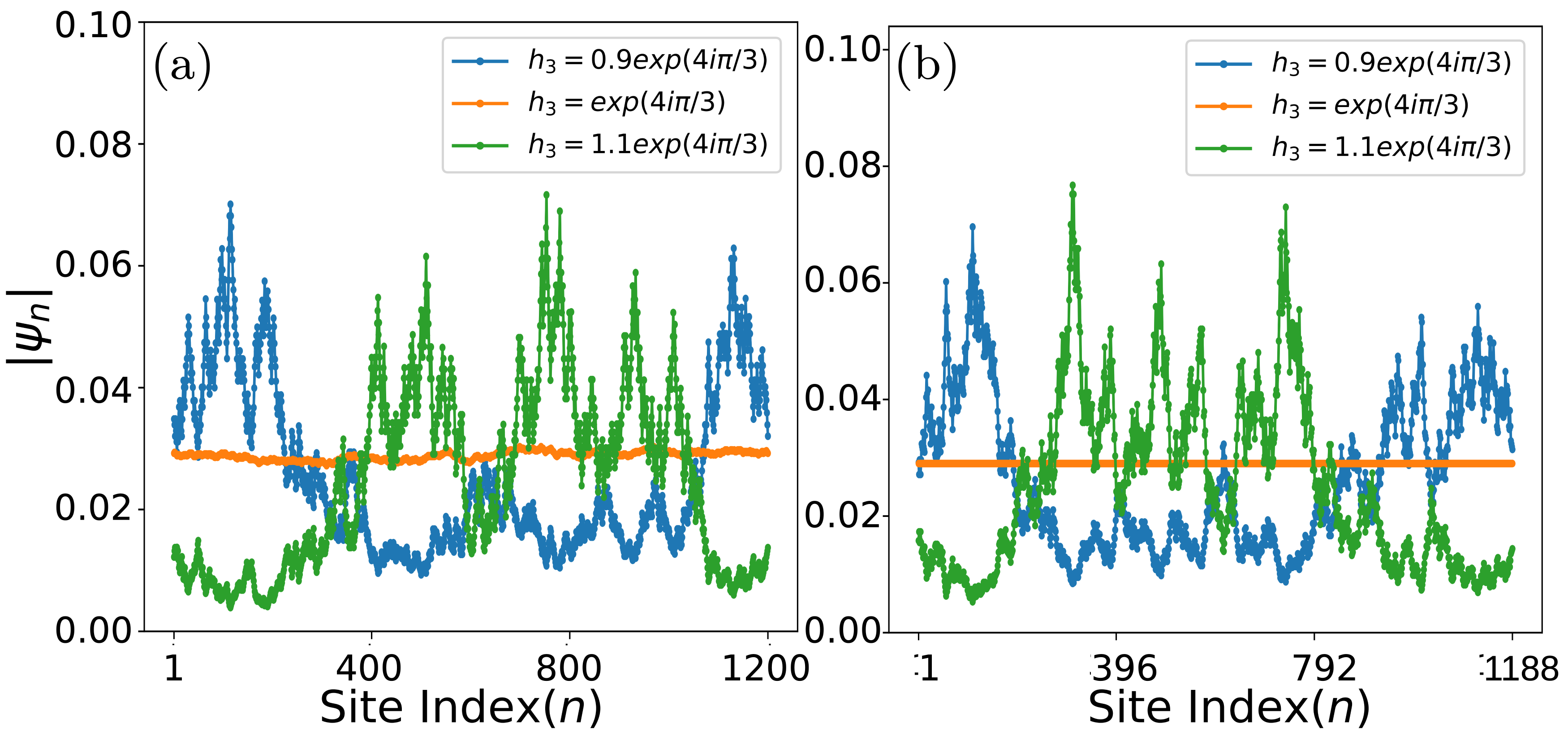}
    \caption{ 
        Absolute value of the eigenstate corresponding to maximum participation number in case of ternary $M=3$ disorder versus the lattice coordinate. Parameters: $t=1$, $h_1 = e^{i\pi/3}$,$h_2 = e^{2i\pi/3}$, panel (a)~$N = 1200$, panel (b)~$N = 1188$.
    }
    \label{fig:ternary_disorder_eigvecs}
\end{figure}

In Fig.~\ref{fig:ternary_disorder_eigvecs} we show the absolute values of the wave functions with largest participation number $P$ for the above three cases, but for two different system sizes. 
For the originally chosen size $N=1200$ the condition \eqref{eq:condition2} is not exactly satisfied. With equal probabilities of the three $h_{1,2,3}$ the total phase shift amounts to $400\frac{7}{3}\pi$ which is violating \eqref{eq:condition2}. 
Indeed, Fig.~\ref{fig:ternary_disorder_eigvecs}(a) shows that while the two cases with $h_3$ off the unit circle result in wave functions with subexponential fluctuations, even the case when $h_3$ is on the unit circle shows slight fluctuations, though of course much smaller and of order $1/N$. 
But when we choose a system size $N=1188$ condition \eqref{eq:condition2} is met exactly since the total phase shift $396\frac{7}{3}\pi$ is a multiple of $2\pi$. 
Consequently, the delocalized wave function for $h_3$ on the unit circle becomes completely and exactly delocalized as shown in Fig.~\ref{fig:ternary_disorder_eigvecs}(b).

It is easy to follow the binary disorder discussion in Sec.~\ref{sec:general_weights} in order to show that changing the weights of the ternary $M=3$ disorder does not affect the results. Extensions to larger finite values of $M$ are straightforward.

\subsection{The limit \texorpdfstring{$M\rightarrow \infty$}{M to infinity}}
\label{sec:the_limit_M_to_infinity}

We consider general disorder in the limit $M \rightarrow \infty$.
The generalized Cassini-oval structure of the spectrum is now lost, but our theorem still holds in the thermodynamic limit (and exactly at finite size when the PBC condition~\eqref{eq:condition2} is satisfied): if all values $\epsilon_n$ are located on a circle in the complex plane with radius $t=1$, then the center of that circle belongs to the spectrum and hosts a completely delocalized eigenstate.
\begin{figure}[htbp]
    \centering    
    \includegraphics[width = 0.99\linewidth]{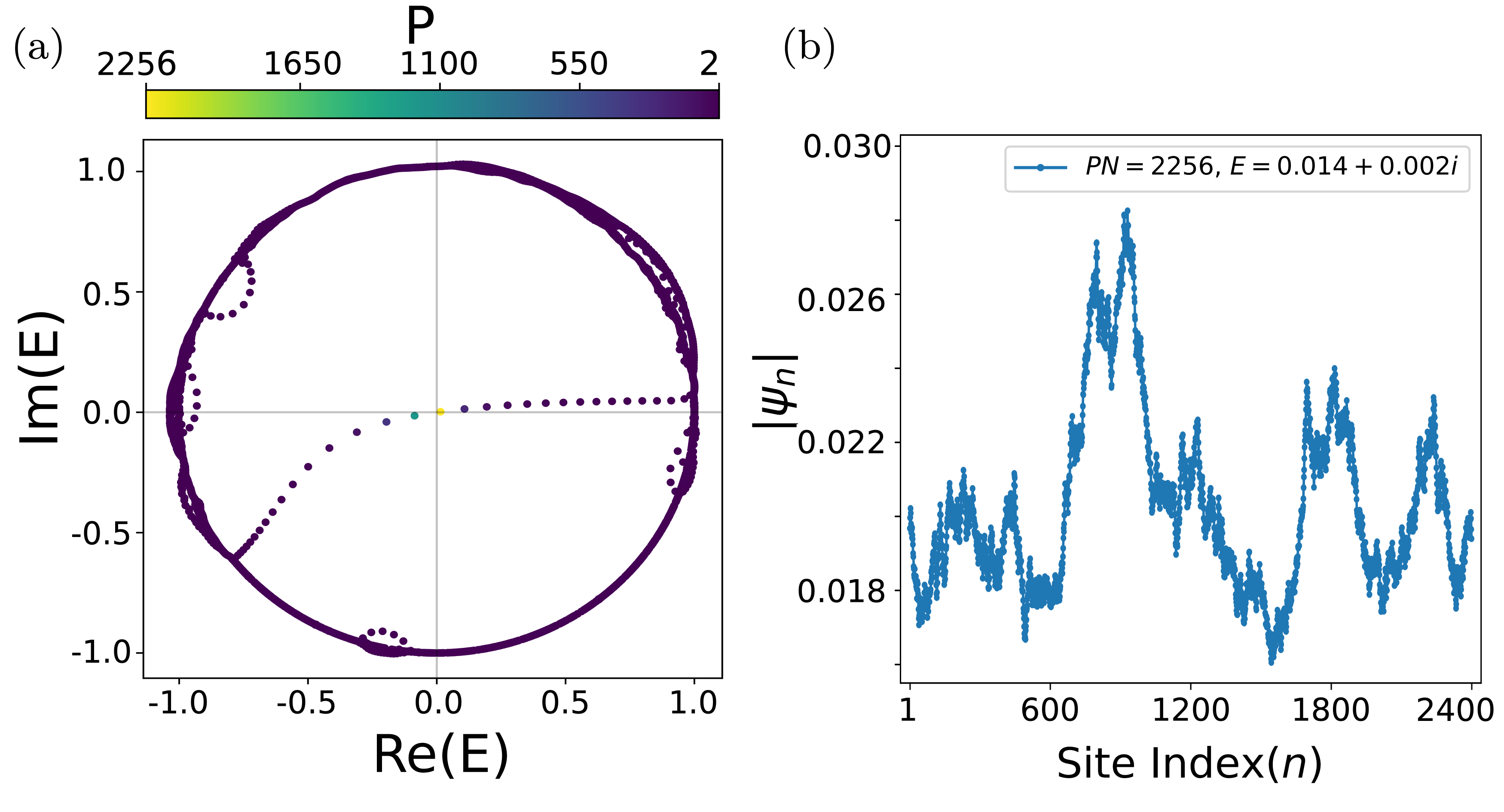}
    \caption{
        Plot of the spectrum and eigenstate corresponding to maximum PN in case of uniform random disorder on a unit circle in complex plane. 
        Fig.~(a): Spectrum in complex plane and PN is shown by color code. Fig.~(b): Eigenstate corresponding to maximum PN. 
        Parameters: $N = 2400$, $t = 1$, $\epsilon_{n} = e^{i\theta_n}$.
    }
    \label{fig:infiniteM_disorder_eigvals_eigvecs}
\end{figure}

In Fig.~\ref{fig:infiniteM_disorder_eigvals_eigvecs}(a) we plot the eigenvalue spectrum for a chain with $N=2400$ and onsite disorder $\epsilon_n={\rm e}^{i\theta_n}$ with uncorrelated phases drawn with equal probability from the interval $0 \leq \theta_n < 2\pi$. All $\epsilon_n$ are located on a circle of unit radius. 
The eigenvalue spectrum forms a main circle-like structure away from the circle center, but in addition features a line connecting two sides of the circle structure and running straight through the circle center. The eigenvalues on this line form a discrete set due to finite system size. 
The eigenstate closest to the circle center has an eigenenergy $E=0.014+0.002i$ and a $P$ value which clearly shows almost complete delocalization. 
Indeed in Fig.~\ref{fig:infiniteM_disorder_eigvals_eigvecs}(b) we plot the absolute values of its wave function along the system size, and show very little fluctuations induced by finite size effects.

\section{Phases for Tailoring}
\label{sec:phases_for_tailoring}

The approach to put base-$M$ disorder on a unit circle in order to get a spectrum with a completely delocalized state in the uHN chain is a particular example of disorder tailoring. 
We will now extend this approach by prescribing the state we want to be part of the eigenvalue problem of some disordered system.

Let us define a target state $\Psi_T$ with $\psi_n = {\rm e}^{i\phi_n}$ by randomly choosing some disorder realization for the phases $\phi_n$, e.g. uncorrelated and randomly distributed over the entire interval $[0,2\pi)$. Define $\Delta\phi_n=\phi_{n+1}-\phi_{n}$. 
For the uHN chain we then obtain the tailored onsite disorder realization $\epsilon_n = E+{\rm e}^{i\Delta\phi_{n}}$ for which the uHN chain will have a spectrum with an eigenvalue $E$ and a completely delocalized state $\Psi_T$. 
This is essentially identical with the results for base-$M$ disorder discussed in Sec.~\ref{sec:base_M_disorder}.

We extend this result to a general HN model 
\begin{equation}
    H = \sum_{n=1}^{N} \epsilon_n \ a_{n}^{\dagger}a_{n} -t_L \ a_{n}^{\dagger}a_{n+1} -t_R \ a_{n+1}^{\dagger}a_{n} 
    \label{eq:ham2}
\end{equation}
and use the above target state to tailor the disorder 
\begin{equation}
    \epsilon_n = E+t_R{\rm e}^{-i\Delta\phi_{n-1}} + t_L {\rm e}^{i\Delta\phi_{n}}\;.
    \label{eq:phasetailoring}
\end{equation}
We can further consider the Hermitian chain $t_L=t_R=1$. 
The above tailoring principle is still correct. However in general the tailored disorder will be complex and make the resulting system non-Hermitian. 
But for special choices of binary phase disorder e.g. $\phi_n =0,\pi$ the resulting real-valued tailored disorder will keep the system Hermitian. 
Then the general eigenstates will be Anderson (exponentially) localized, but the eigenstate at the tailored eigenvalue $E$ will be completely delocalized, due to the fact that the tailored disorder has nonzero short range correlations.

Let us fix the tailored state energy $E=0$. Define $s_n={\rm e}^{i\phi_n}$. 
With $\phi_n$ taking uncorrelated values $0$ or $\pi$ the random numbers $s_n$ take uncorrelated random numbers $\pm 1$ with equal probabilities.
Then the tailoring Eq.~\eqref{eq:phasetailoring} turns into
\begin{equation}
    \epsilon_n = \kappa_n+\kappa_{n+1}\;,\; \kappa_n=s_{n-1}s_n\;.
    \label{eq:hermitianphasetailoring}
\end{equation}
which is used for the eigenvalue equation
\begin{equation}
    E\psi_n = \epsilon_n \psi_n - \psi_{n-1}-\psi_{n+1}\;.
    \label{eq:hermitianev}
\end{equation}
The disorder field $\kappa_n$ is uncorrelated since $s_n^2=1$ and $ \langle \kappa_n \kappa_{n+1} \rangle = \langle s_{n-1}s_n^2s_{n+1}\rangle= \langle s_{n-1} s_{n+1}\rangle = 0 $.
Therefore the diagonal disorder $\epsilon_n$ has a built in nearest neighbor correlations, but no next-to-nearest neighbor correlations.

\begin{figure}[htbp]
    \centering
    \includegraphics[width=0.99\linewidth]{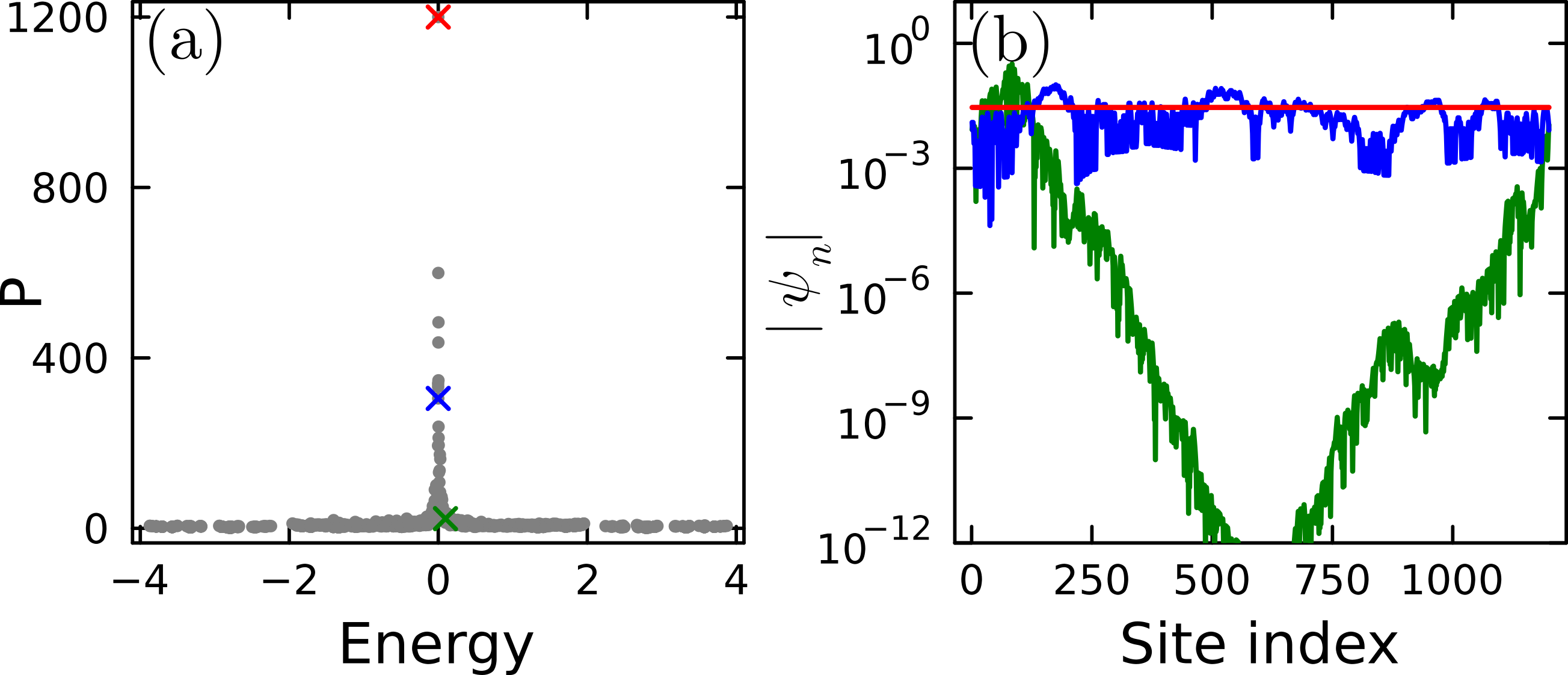}
    \caption{
        Participation numbers and representative eigenstates of the Hermitian tailored-disorder chain. 
        Parameters are $N=1200$, $\epsilon_n=\kappa_n+\kappa_{n+1}$, $\kappa_n=s_{n-1}s_n$, and $s_n=\pm1$. 
        (a) Participation number $P$ versus eigenenergy $E$ for one disorder realization. 
        The exactly delocalized state at $E=0$ has $P=N$, while the states away from zero are localized. The three highlighted points correspond to the eigenstates plotted in panel (b). 
        (b) Absolute values of the three representative eigenstates: red, $E=0$; blue, $E = 0.0006$; and green, $E = 0.1$. 
        The $E=0$ eigenstate has a perfectly uniform amplitude, whereas spatial fluctuations and localization develop with increasing distance from the tailored energy. 
    }
    \label{fig:tailor_eigvals_eigvecs}
\end{figure}

\begin{figure}[htbp]
\centering
    \includegraphics[width = 0.9\linewidth]{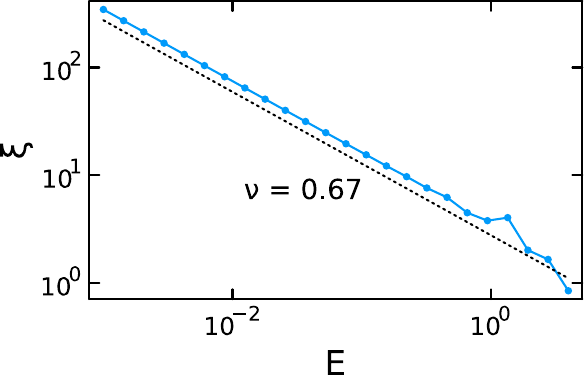}
    \caption{ 
        Localization length as a function of energy for the Hermitian Anderson chain with disorder~\eqref{eq:hermitianphasetailoring}. 
        Parameters: $\epsilon_{n} = \kappa_{n}+\kappa_{n+1}$, $\kappa_{n} = s_{n-1}s_{n}$ , $s_{n} = \pm 1$ with equal probability. 
        Blue curve and symbols: numerical results, black dashed line: power law fit.
        A clear power law divergence is observed with exponent close to $\nu \approx 2/3$.
    }
    \label{fig:localizationlengthdivergence}
\end{figure}

We numerically test our prediction. 
The participation number versus the eigenvalue spectrum for $N=1200$ is shown in Fig~\ref{fig:tailor_eigvals_eigvecs}(a). 
Almost all eigenstates show very small values of $P$, but a diverging peak at $E=0$ is clearly observed, with the value of $P = 1200$.
In Fig.~\ref{fig:tailor_eigvals_eigvecs}(b) we plot the absolute values of the wave function of the eigenstate at $E=0$ versus space. 
Ideal delocalization is observed for the $E = 0$ eigenstate up to machine precision (red). 
To further quantify our observation, we compute the localization length of all eigenstates using the transfer matrix approach (which in this context turns into a transfer scalar approach). 
We note that the eigenvalue spectrum is confined to the real interval $[-4,4]$ since $\epsilon_n$ can only take values $-2,0,+2$ and the hopping part of the Hamiltonian contributes additively with a spectrum confined between $-2$ and $+2$. 
Next we define $R_n=\frac{\psi_n}{\psi_{n-1}}$ 
and obtain the recursion equation
\begin{equation}
    R_{n+1} = \epsilon_n - E - 1/R_n
    \label{eq:transfermatrix}
\end{equation}
which can be initialized with any value of $R_1$ e.g. $R_1=1$.
For each allowed value of $E$ we compute the average of $\xi^{-1}(E)=\lim_{N_{tr} \rightarrow \infty}\frac{1}{N_{tr}} \sum_{n=1}^{N_{tr}}\ln |R_n|$ with $\xi(E)$ being the localization length describing the asymptotic exponential decay of the eigenstate $|\psi_n| \sim {\rm e}^{-n/\xi}$ for large $n$. 
In practice we stop the averaging for a sufficiently large number of iterations $N_{tr}$ making sure that the averaging converged. 
Then plot the localization length versus energy in Fig.~\ref{fig:localizationlengthdivergence}. 
We observe algebraic divergence $\xi \sim |E|^{-\nu}$ with the exponent $\nu \approx 2/3$. 

To derive the exponent $2/3$ for the divergence of the localization length, we write $\psi_n=s_nu_n$ and define
\begin{align}
    J_n=\kappa_n(u_n-u_{n-1}).
    \label{eq:Jn}
\end{align}
The eigenvalue problem [Eqs.~\eqref{eq:hermitianev} and
\eqref{eq:hermitianphasetailoring}] then transforms to
\begin{align}
    Eu_n=J_n-J_{n+1}.
    \label{eq:tailor_ev_mapping}
\end{align}
For $E\ne0$, we substitute $u_n=(J_n-J_{n+1})/E$ and
$u_{n-1}=(J_{n-1}-J_n)/E$ into Eq.~\eqref{eq:Jn}. Using $\kappa_n^2=1$, we obtain
\begin{equation}
    J_{n+1}+J_{n-1}+E\kappa_nJ_n=2J_n.
    \label{eq:newev}
\end{equation}

We therefore mapped the original eigenvalue problem with diagonal short range correlated disorder at a finetuned fixed strength~(\ref{eq:hermitianphasetailoring},\ref{eq:hermitianev}) and a built-in completely delocalized state in the center of the spectrum to an eigenvalue problem for a state at fixed eigenenergy $\tilde{E}=2$ and with a purely uncorrelated diagonal disorder strength $\tilde{\epsilon}_n = E \kappa_n$ which is linear in the original eigenenergy $E$. 
Therefore the eigenstate for $E\rightarrow 0$ must become delocalized. 
Note also that the fixed eigenenergy $\tilde{E}=2$ corresponds to the spectrum band edge of \eqref{eq:newev} for the clean limit in the absence of any disorder.
This maps the original tailored model to a one-dimensional Anderson chain at the clean band edge $2$, with uncorrelated on-site disorder of
amplitude $|E|$. 
The known band-edge scaling of the Lyapunov exponent, $\gamma\propto W^{2/3}$~\cite{derrida1984lyapounov}, therefore gives $\xi(E)=\gamma^{-1}\propto |E|^{-2/3}$, in full agreement with our numerical observations.

Since the delocalized state, purified from its random phases, is simply $u_n=1$ (up to normalization), it is tempting to assume that a form of (generalized) Galilean symmetry is responsible for that delocalization, similar to the random mass chain \cite{matsuda1970localization,ishii1973localization} discussed in Sec.~\ref{sec:introduction}. 
Indeed, (\ref{eq:hermitianev}) with $\psi_n=s_nu_n$ turns into $Eu_n=\kappa_n(u_n-u_{n-1})+\kappa_{n+1}(u_n-n_{n+1})$ which is generated by a classical Hamiltonian of masses connected by harmonic springs with random spring constants $\pm1$: 
\begin{equation}
    H=\sum_n \frac{p_n^2}{2}+\frac{\kappa_n}{2}(x_n-x_{n-1})^2\;.
    \label{eq:ham_coupled_eom}
\end{equation}
Here $p_n$ is a momentum of a point mass $m=1$, and $x_n$ is its canonically conjugated displacement.
The ansatz $x_n = u_n {\rm e}^{i\omega t}$, together with the Hamiltonian equations of motion~\eqref{eq:ham_coupled_eom}, results in the above eigenvalue problem for $u_n$ with $E=\omega^2$. 
Disregarding the aspect that the spring Hamiltonian is inherently unstable, we note that it possesses Galilean symmetry. 
Our tailoring method is therefore generating models with predefined symmetries. 
It is straightforward to apply the above considerations to our uHN results in Sec.~\ref{sec:disordered_hatano_nelson_models}, which results in a similar generalized Galilean symmetry at work. 

The above phase tailoring can be easily extended to any higher lattice dimension. 
The reason is simply that for each local site $n$ we will have one equation to be satisfied, and we have exactly one variable $\epsilon_n$ which we tailor to satisfy that equation. 
The Galilean symmetry considerations carry over in the same way.

\section{Instead of conclusions: extensions to more}
\label{sec:instead_of_conclusions_extensions_to_more}

\subsection{Hoppings for tailoring}
\label{sec:hoppings_for_tailoring}

We can easily extend the tailoring method by adding any disorder realization of the hoppings. 
We can even increase the hopping range, and introduce randomness again. We may keep the system Hermitian by keeping the target state phases $\phi_n=0,\pi$, add the Hermitian hopping disorder, and compute the tailored onsite disorder to satisfy a homogeneous target state as a function of the input hopping and phase disorder fields. 
Needless to say, that will work on any lattice with any lattice dimension, Hermitian or non-Hermitian. 
Non-Hermiticity can be induced from scratch by choosing non-Hermitian hopping fields, and/or phase fields of the target state which will induce non-Hermiticity as well. We could as well design a disorder phase field which ensures Hermiticity everywhere in the system except for a single non-Hermitian impurity site (as studied e.g. in general non-tailored cases in Ref.~\cite{molignini2023anomalous}) or even in various regularly or randomly distributed clusters of non-Hermiticity.

\subsection{Changing the target states}
\label{sec:changing_the_target_states}

However, we can also in addition change the target state into anything we like, e.g. exponentially decaying in one direction, being a homogeneous state in the other direction, or decaying algebraically in the other direction, or anything else we prefer, with even more freedom we can execute in higher lattice dimensions. 
After defining the target state, and the optional hopping fields, we proceed with computing the tailored disorder on each site $n$ which will support that target state as an eigenstate.

\subsection{Many body networks}
\label{sec:many_body_networks}

Finally we note that the tailoring method can be readily used for specific network types used to describe many body physics, in particular Cayley trees, Bethe lattices \cite{prado2026anderson} and random regular graphs \cite{tikhonov2021anderson}. Similar to the example in Sec.~\ref{sec:phases_for_tailoring}, we will first fix the target energy and target wave function. Then we will configure the hopping part of the network. Finally we compute (tailor) the on site energy $\epsilon_n$ for each site to satisfy the eigenvalue equation. The outcome is a Cayley tree, Bethe lattice, or random regular graph with a tailored short range correlated disorder which will support a predefined eigenstate (e.g. one with constant amplitude and random phases) at a predefined energy. Of course one can again modify the target state, as well as add randomness to the hopping fields on these networks.

\begin{acknowledgments}
    We thank Oleg Evnin, Keith Slevin, Miguel Ortuno,  Napat Poovuttikul and Alexei Andreanov for useful discussions. 
    The authors acknowledge financial support from the Institute for Basic Science (IBS) in the Republic of Korea through Project No. IBS-R024-D1 and IBS-R041-D1-2026-a00.
    SG acknowledges financial support from Indian Institute of Technology-Kanpur, India through Institute Post-Doctoral Fellowship (PF. No. PDF604). 
\end{acknowledgments}

\bibliography{used.bib}

\end{document}